\documentclass{article}\usepackage{epsfig,latexsym,subequation,graphics}
\def\circa#1{\,\raise.3ex\hbox{$#1$\kern-.75em\lower1ex\hbox{$\sim$}}\,}
\newcommand{\ov}{{\cal O}}  \newcommand{\f}{{\phi}}
\newcommand{\Ttot}{{\bf T}} 

 \newcommand \bra {\langle}
\newcommand \ket {\rangle} \newcommand{\be}{\begin{equation}}
\newcommand{\ee}{\end{equation}} \newcommand{\ben}{\begin{displaymath}}
\newcommand{\een}{\end{displaymath}} \newcommand{\ba}{\begin{eqnarray}}
\newcommand{\ea}{\end{eqnarray}} \newcommand{\ban}{\begin{eqnarray*}}
\newcommand{\ean}{\end{eqnarray*}} \newcommand{\cro}{\dagger}
 \newcommand{\de}{\partial}

\newcommand{\Phivet}{\mbox{\boldmath $\Phi$}}
\newcommand{\kt}{\mbox{$k$}_{\perp }}

\newcommand{\Ytot}{{\rm\bf Y}}

\newcommand{\effe}[1]
{\mathop{{\vspace{3pt}\cal F}}\limits^{}_{\hskip-1mm \scriptscriptstyle #1}}

\newcommand{\ff}[2]{\mathop{{\vspace{3pt}f}}\limits^{}_{\hskip-1mm
  \scriptscriptstyle #1}{\hskip-1mm \scriptstyle #2}}

\newcommand{\F}[2]
{\mathop{{\vspace{3pt}\cal F}}\limits^{#1}_{\hskip-1mm \scriptscriptstyle #2}}
\newcommand{\der}{-\frac{\de}{\de t}}
\newcommand{\overlap}[1]
{\mathop{{\vspace{3pt}\cal O}}\limits^{}_{\scriptscriptstyle #1}}
 
\begin{document}

\vspace{1.cm}

{\centering

{\Large\bf Electroweak Evolution Equations }

\vspace{1.cm}

{\bf \large Paolo Ciafaloni}

{\it INFN - Sezione di Lecce, \\Via per Arnesano, I-73100 Lecce, Italy \\
E-mail: paolo.ciafaloni@le.infn.it}
\vspace{0.4cm}

{\bf \large Denis Comelli}

{\it INFN - Sezione di Ferrara, \\Via Paradiso 12, I-35131 Ferrara, Italy\\
E-mail: comelli@fe.infn.it}

}

\vspace{0.3cm}

\begin{abstract}
Enlarging a previous analysis, where only fermions
and transverse gauge bosons were taken into account, we write down infrared-collinear
evolution equations for the Standard Model of electroweak interactions
computing the full set of splitting functions.  
Due to the presence of double logs which are characteristic of electroweak 
interactions (Bloch-Nordsieck violation), new infrared singular splitting functions have
to be introduced. 
We also include corrections related to the third generation Yukawa couplings.
\end{abstract}

\section{\bf Introduction}

Energy-growing electroweak corrections in the Standard Model have received
recently a lot of 
 attention in the literature, being relevant for LHC physics \cite{acco},
for Next generation of Linear Colliders (NLCs) \cite{nlc}
 and for ultrahigh energy cosmic
rays \cite{berez}. The presence of double logs   
($\log^2\frac{s}{M^2}$ where $M$ is the weak scale)
in one loop electroweak corrections
has been noticed in \cite{first}.
One loop  effects are
typically of the order of 10-20 \%  at the  energy  scale of 1
TeV,  so that the subject of higher orders and/or resummation of large
logarithms has to be addressed. 
After the observation  \cite{cc} that double and single logs
that appear in the 1 loop expressions are tied to the infrared structure of
the theory, all order resummation has been considered at various levels:
Leading Log (LL) \cite{LL}, Next to Leading Log (NLL) \cite{NLL} and so on.
Moreover many fixed order analyses at the one \cite{oneloop} and two loop
\cite{twoloop} level have been performed.

Collinear evolution equations 
are written in general with the purpose of resuming large contributions of
logarithmic type by factorizing collinear singularities. This is done by
separating a ``soft'' scale contribution that includes collinear logarithms
 and a ``hard'' scale contribution which is
free of logarithms and therefore perturbative. However in the electroweak
sector at energies $\sqrt{s}$ much higher than the weak scale 
$M\sim M_Z\sim M_W\approx 100$ GeV, logarithms of both collinear {\sl and}
infrared origin appear, even in fully inclusive quantities \cite{3p1}. 
In a previous paper \cite{col} we have written 
infrared evolution equations in the limit of vanishing U(1) coupling and
considering only fermions and transverse gauge bosons, showing that the
presence of both infrared and collinear singularities can be tackled with
by introducing new, infrared singular, splitting functions.
 The aim of this work is to complete the analysis by writing down infrared evolution 
evolution equations in the full SU(2)$\otimes$ U(1) 
electroweak sector of the Standard Model, including 
left and right handed fermions, transverse and 
longitudinal bosons degrees of freedom.
Since the third family Yukawa couplings are non negligible, we include their
contribution; on the other hand we neglect family mixing effects related to
the Cabibbo  Kobayashi Maskawa matrix.
 A complete analysis of collinear
singularities in the Standard Model will also have to include
QED evolution  equations \cite{QED} for transverse momenta of the emitted particles lower
than the weak scale $M$, and QCD DGLAP \cite{DGLAP} evolution equations.

The analysis of mass singularities in a spontaneously broken gauge theory
like the electroweak sector of the Standard Model has many interesting
features.   To begin with, initial states like electrons and
protons carry nonabelian (isospin) charges; this
feature causes the very existence of double logs i.e. the lack of cancellations
of virtual corrections with real emission in inclusive observables \cite{3p1}.
Secondly, initial states that are mass eigenstates are not necessarily
gauge eigenstates; this causes some interesting mixing phenomena 
analyzed in \cite{full,lon,abelian}. Technically speaking, a complication
is due to the fact that the gauge theory Ward Identities are broken by Goldstone boson
contributions;  however these contributions 
are proportional to the symmetry breaking scale and are therefore
expected to be suppressed to a certain extent in the high energy limit.
These contributions have been shown to be negligible at the one loop level 
\cite{col}; we assume this to be true at higher orders as well, although we
lack a formal proof  at the moment. Finally, collinear factorization has
been proved at the one loop level \cite{Denner:2000jv}; we assume it to be
valid to all orders.  Ultraviolet logarithms leading to running coupling effects are neglected throughout the
paper.

The main results of this paper are contained in 
eqs. (\ref{fulleqs}-\ref{endeq}), that represent  
the   SU(2)$\otimes$U(1) electroweak evolution equations for the full SM spectrum.
Namely,  eqs. (\ref{fulleqs}) are written for matricial structure functions 
$\F{i}{j}\!_{\alpha'\beta'}^{\alpha\beta}$ 
in  isospin space (see fig. 1). 
In order to make  these equations useful for practical purposes,
 we have to write them for corresponding scalar quantities.
 This is done by exploiting the (recovered) isospin and CP symmetry,
 i.e. by projecting the equations on states of given quantum numbers,
 diagonalizing, in such a way, the system  in block structures  characterized by conserved quantum numbers.

The full set of evolution equations are computed working with gauge eigenstates
which simplify systematically the evaluation in the high energy regime.
We do this by giving in the Appendix  a complete classification of possible 
asymptotic states according to their isospin and CP properties.
Then, scalar equations are obtained by a method we introduce here, which
consists in performing appropriate traces with respect to isospin leg indices.
The final equations for scalar structure functions $\ff{i}{\scriptstyle(T,Y;CP)}$ 
are given in   (\ref{inizio}-\ref{fine}). 

Since the overall procedure is quite complicated, in section two
we discuss the simple case of left fermions in the initial state and
 we show  how the  block diagonalization procedure 
helps for  the numerical and practical  evaluation of the evolution equations.

\section{A working example: left fermions in the $g'\to 0$ limit}

In this section we consider lepton initiated Drell Yan process
of type $e^+(p_1)\;e(p_2)\rightarrow q(k_1) \bar{q}(k_2)  +X$\footnote{note
that the process considered, like all others in this paper, is fully
inclusive, meaning also $W,Z$ radiation is included.} 
where $s=2 p_1 \cdot p_2$ is the total invariant mass and $Q^2=2 k_1 \cdot k_2$
 is the hard scale. We consider double log corrections in relation to the
 SU(2) electroweak gauge group, i.e. we work in the limit where the U(1)
 coupling   $g'$ is zero. This process has been analyzed in \cite{strong,full}, 
to which we refer for details; we consider it here for convenience and 
in order to establish our notations.
The general formalism used to study electroweak evolution equations for
inclusive observables has been set up in \cite{col}; we summarize it here
briefly. To begin with, by arguments of unitarity, final state
radiation can be neglected when considering inclusive cross sections
\cite{strong}. Then we are led to consider the dressing
of the overlap matrix $\overlap{}\!_{\alpha\beta}^{\alpha'\beta'}= \bra \beta\beta'|
S^+\;S|\alpha\alpha' \ket$, $S$ being the $S$-matrix,
 where only initial states  indices appear explicitly (see fig. \ref{f1}). 

At the leading level, all order resummation in the soft-collinear region 
is obtained by a simple expression  that involves
the t-channel total isospin $T$ that couples indices  $\alpha,\beta$:
\be\label{llll}
\ov^H\to \ov^{resummed}= e^{-\frac{\alpha_W}{4\pi}[T(T+1)]
\log^2\frac{s}{M_w^2}}
\;\;\ov^H
\ee
$\alpha_W$ being the weak coupling, $\ov_H$ the hard overlap matrix written in
terms of the tree level $S$-matrix and  $M_w\sim M_z$.

At subleading order, the dressing  by soft and/or collinear radiation
is described at all orders by infrared evolution equations, that are 
$T$-diagonal as far as fermions and transverse gauge bosons 
are concerned \cite{col}. In order to write down the evolution equations
for the case of initial left fermions, we first consider one loop corrections.
At the one loop level, virtual and
real corrections in NLL approximation can be
written as:

\begin{equation}\label{cicciotto}
           \begin{eqalign}   
\delta\overlap{L}\!_{\alpha\beta}&=
\frac{\alpha_W}{2\pi}\int_{M^2}^s\frac{d\kt^2}{\kt^2}
\int_0^{1}
\frac{dz}{z}\left\{
P^R_{ff}(z)\;\theta(1-z-\frac{\kt}{\sqrt{s}})
\;t^A_{\beta\beta'} \;t^A_{\alpha'\alpha}\; \overlap{L}\!^H\!\!\!_{\alpha'\beta'}\!(zp)\;+
\right.\\
&\left.
P^R_{gf}(z)\;[t^B \,t^A]_{\beta\alpha} \;\overlap{g}\!_{AB}^H(zp)
+ C_f \;
P^V_{ff}(z,\frac{\kt}{\sqrt{s}})\; \overlap{L}\!_{\alpha\beta}^H(p)
\right\}
\end{eqalign}\ee
where
$ P^R_{ff}(z),\,P^R_{gf}(z)$ and $P^V_{ff}(z,\kt) $ 
are defined in the Appendix. The indices   below the overlap matrix label
the kind of particle:
$L=$ Left fermion and $g=$ gauge boson ;
 indices $\alpha,\beta$ refer to the isospin index  ($\alpha=1$ corresponds
 to $\nu$, $\alpha=2$ to $ e$) of the
 lower legs while upper legs
indices are omitted. 

\begin{figure}
      \centering
      \includegraphics[height=80mm]
                  {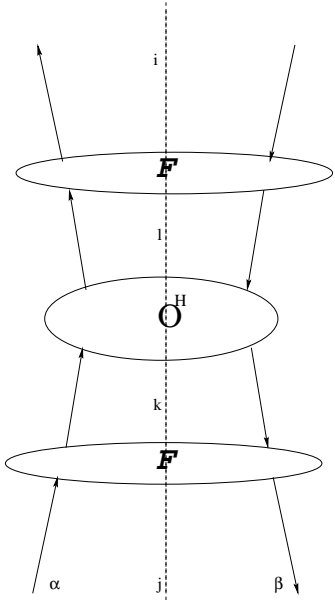}
     
 \caption{\label{f1}
Graphical picture of the factorization formula \ref{fact}.)
 }  
\end{figure}

The one loop formula (\ref{cicciotto}) is consistent
 with a general factorization formula
of type (see fig. \ref{f1}))
\be\label{fact}
\mathop{{\vspace{3pt}\cal O}}\limits^{ i}_{j}
(p_1,p_2;k_1,k_2)=\int \frac{dz_1}{z_1}
\frac{dz_2}{z_2}
\sum_{k,l}^{L,R,g}
\mathop{{\vspace{3pt}\cal F}}\limits^{k}_{j}(z_1;s,M^2)
\mathop{{\vspace{3pt}\cal O}}\limits^{l}_{k}\!^H
(z_1 p_1,z_2 p_2;k_1,k_2) \F{i}{l}(z_2;s,M^2)
\ee
where $i,j$ label the  kind of particle ($L$=left fermion,
$g$=gauge  boson),
 and  where
isospin flavor indices in the overlap function $\ov$ and 
structure function ${\cal F}$ are 
understood.

If the factorization formula (\ref{fact}) is assumed to be valid 
at higher orders as well,
the structure functions will satisfy evolution equations with respect to 
an infrared-collinear  cutoff
$\mu$ parameterizing the lowest value of $\kt$, as follows ($t=\log\mu^2$):
\be\label{kpil}
\der\F{\scriptscriptstyle i}{j}\!_{\alpha\beta}=
\frac{\alpha_W}{2\pi}\left\{
C_f \F{\scriptscriptstyle i}{j}\!_{\alpha\beta} \otimes P^V_{ff}
+[t^C\F{\scriptscriptstyle i}{j}\,^t\; t^C]_{\alpha\beta}\otimes P_{ff}^R
+[t^B\;t^A]_{\beta\alpha}\; \F{\scriptscriptstyle i}{g}\!_{AB}
\otimes P^R_{gf}
\right\}
\ee
In these equations $t^A$ denote the isospin matrices
 in the fundamental
representation and $\F{i}{j}\!{\scriptscriptstyle\alpha\beta}$ 
denotes the distribution of a
particle $i$ (whose isospin indices are omitted)
inside particle $j$ (with isospin
leg indices ${\alpha,\beta}$).
$\F{\scriptscriptstyle i}{j}\,^t$ is the transpose
matrix $\F{i}{j}\!{\scriptscriptstyle\beta\alpha} $. Furthermore, we have defined
the convolution
$
[f\otimes P](x)\equiv\int_x^1 P(z)f(\frac{x}{z})\frac{dz}{z}
$; the relevant splitting functions are given in Appendix.
Since the index $i$ is always kept fixed in (\ref{kpil}), 
we will omit it from now on,
with the understanding that, for instance, 
$\mathop{{\vspace{3pt}\cal F}}\limits^{}_{j}$ collectively
denotes 
all $\mathop{{\vspace{3pt}\cal F}}\limits^{i}_{j}$ with any 
value of $i$. 

Eqn. (\ref{kpil}) is a matricial   evolution equation; in order to make it
useful we can write the corresponding scalar equations. We do this by
exploiting the $SU(2)_L$ symmetry which allows to classify the states
according to their isospin quantum numbers. We couple the lower legs
$\alpha, \beta$ in fig. 1 obtaining the t-channel isospin eigenstates:
\be
|{\bf T}=0\ket=\frac{1}{\sqrt{2}}(|\nu\nu^*\ket+|ee^*\ket)\qquad
|{\bf T}=1\ket=\frac{1}{\sqrt{2}}(|\nu\nu^*\ket-|ee^*\ket)
\ee
which have $T_L^3=0$ since cross sections always have a given particle on
leg $\alpha$ and its own antiparticle on leg $\beta$. We now project the
structure operators ${\cal F}$ on these states, omitting the upper leg
indices:
\be\label{xcqef9}
\ff{L}{(0)}=\frac{\bra \nu\nu^*+ee^*|\F{}{L}|\ket}{2}=
\frac{\effe{L}\!_{\nu\nu}+\effe{L}\!_{ee} }{2}  
=\frac{1}{2} {\rm Tr}\left[\F{}{L}\right]
\qquad \ff{L}{(1)}=\frac{\bra \nu\nu^*-ee^*|\F{}{L}|\ket}{2}=
\frac{\effe{L}\!_{\nu\nu}-\effe{L}\!_{ee}}{2}=
{\rm Tr}\left[t^3\F{}{L}\right]
\ee
Last step in eqs. (\ref{xcqef9}) represents a
 convenient way  to extract the scalar coefficients 
$ \mathop{{\vspace{3pt}f}}\limits^{}_{j}\!^{(T)}$ from 
 $\mathop{{\vspace{3pt}\cal F}}\limits^{}_{j}$; namely,   
by taking  appropriate traces with respect to the
soft leg $j$. For instance $\ff{L}{\scriptstyle(0)}$ 
corresponds to $\frac{1}{2}(
\mathop{{\vspace{3pt}\cal F}}\limits_{L}\!_{ee}+
\mathop{{\vspace{3pt}\cal F}}\limits_{L}\!_{\nu\nu})$  \cite{col} and can
be obtained by 
$\mbox{Tr}_j[\mathop{{\vspace{3pt}\cal F}}\limits^{}_{j}]$; 
here and in the following 
 the trace is taken
 with respect to the indices of the soft lower scale leg
$j$ .
Notice that since gauge and mass eigenstates do not necessarily coincide,
we have to introduce ``mixed legs'' with particles belonging to different
gauge representations on leg $\alpha$ and $\beta$ (more about this point, in section 4). We label these cases by
$i=LR$ for the mixed left/right fermion leg, 
$i=B3$ for the mixed $W_3-B$ gauge bosons and
$i=h3$ for the Higgs sector case. These mixing phenomena are interesting by
themselves and have been considered in \cite{full,lon,abelian,transverse} at double log level.

Projecting eq. (\ref{kpil}) for instance 
on the $T=0$ component we obtain:
\be
\der\mbox{Tr}[\F{}{f}]=
\frac{\alpha_W}{2\pi}\left\{
C_f \mbox{Tr}[\F{}{f}] \otimes P^V_{ff}
+\mbox{Tr}[t^C\F{}{f}\,^t\;t^C]\otimes P_{ff}^R
+\mbox{Tr}[t^B\,t^A]\;\F{}{g}\!\!_{AB}\;\otimes P^R_{gf}
\right\}
\ee
where the traces are taken, here and in the following,
with respect to the soft leg indices. This 
gives:
\be\label{andra}
\der \ff{L}{(0)}=
\frac{\alpha_W}{2 \pi}\;\left(
\frac{3}{4}\; \ff{L}{(0)}\otimes \left( P^V_{ff}+P^R_{ff} \right)+
 \frac{3}{4}\, \ff{g}{(0)}\otimes P^R_{gf}\right)
\ee
after taking into account that 
$\mbox{Tr}[t^B\;t^A]\,\F{}{g}\!\!_{AB}=\frac{1}{2}\sum_A\F{}{g}\!_{AA}=
\frac{1}{2}\mbox{Tr}[\F{}{g}]$ and
that $\ff{g}{(0)}=\frac{1}{3}\mbox{Tr}[\F{}{g}]$.

After this short introduction about the use of our projection technique, we now
briefly show the utility  to project on states of definite t-channel quantum number
instead to have evolution equations for single particles.
Following ref.\cite{col} we can define the
  projections with  definite value of the
t-channel isospin $T=0,1,2$ for fermion and gauge boson
\be\label{tt}
\ff{L}{(1)}={\rm Tr}\{t^3\F{}{f}\}=\frac{f_\nu-f_e}{2},\quad
\ff{g}{(0)}=\frac{f_++f_3+f_-}{3},\quad
\ff{g}{(1)}=\frac{f_+-f_-}{2},\quad
\ff{g}{(2)}=\frac{f_++f_--2f_3}{6}
\ee
where 
$f_e=\effe{L}\!_{ee},\;f_\nu=\effe{L}\!_{\nu\nu},
\;f_+=\effe{g}\!_{+-},\;f_-=\effe{g}\!_{-+},\;f_3=\effe{g}\!_{33}$.

We can evaluate the evolution eqs. for the transverse gauge bosons system
 following the same line of the fermionic ones (see ref.\cite{col} or next chapter).
Then, using our  projection technique described before we get the
 scalar equations 
with definite $T$ values;
 precisely there are five equations coupled in three subsets characterized by the 
  $T=0,1,2$ values:
\ba\label{system}
{\rm 5\;factorized\,eqs. } \begin{cases}
 \begin{cases}
\der \ff{L}{(0)}=\frac{\alpha_W}{2\pi}\left(
\frac{3}{4}\ff{L}{(0)}\otimes(P_{ff}^R+P^V_{ff})+\frac{3}{4}\ff{g}{(0)}\otimes
P^R_{gf}\right)
\\
\der \ff{g}{(0)}=\frac{\alpha_W}{2\pi}\left(
2 \ff{g}{(0)}\otimes(P_{gg}^R+P^V_{gg})+\frac{1}{2}
(\ff{L}{(0)}+\ff{\bar L}{(0)})\otimes
P^R_{fg}\right)
\\
\end{cases}\\
\begin{cases}
\der \ff{L}{(1)}=\frac{\alpha_W}{2\pi}\left(
\ff{L}{(1)}\otimes P^V_{ff}-
\frac{1}{4}\ff{L}{(1)}\otimes(P_{ff}^R+P^V_{ff})+\frac{1}{2}\ff{g}{(1)}\otimes
P^R_{gf}\right)
\\
\der \ff{g}{(1)}=\frac{\alpha_W}{2\pi}\left(
\ff{g}{(1)}\otimes P^V_{gg}+\ff{g}{(1)}\otimes(P_{gg}^R+P^V_{gg})+\frac{1}{2}(\ff{L}{(1)}+\ff{\bar L}{(1)})
\otimes   
P^R_{fg}\right)
\end{cases}\\
\begin{cases}
\der \ff{g}{(2)}=\frac{\alpha_W}{2\pi}\left(
3\ff{g}{(2)}\otimes P^V_{gg}-\ff{g}{(2)}\otimes(P_{gg}^R+P^V_{gg})\right)
\end{cases}
\end{cases}
\ea
with similar equations holding for $\ff{\bar L}{(T)}$.
\begin{figure}
      \centering
      \includegraphics[height=80mm]
                  {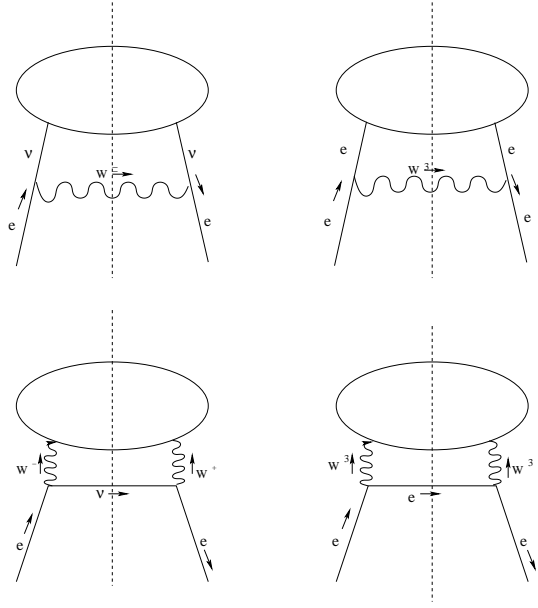}
     
 \caption{\label{en1}
Graphical picture showing the evolution  of   the single electron component, see eqs.
(\ref{system1}).
 }  
\end{figure}

Eqs.(\ref{system}) can be converted in  evolution equations for  
the single components (see fig.\ref{en1}):
\ba\label{system1}
{\rm 5\; eqs}
\begin{cases}
-\frac{4\pi}{\alpha_W}\frac{\de f_{\nu}}{\de t}=
f_\nu\otimes(3P^V_{ff}+P^R_{ff})+2f_e\otimes P^R_{ff}
+2f_+\otimes P^R_{gf}+f_3\otimes P^R_{gf}
\\
-\frac{4\pi}{\alpha_W}\frac{\de f_{e}}{\de t}=
f_e\otimes(3P^V_{ff}+P^R_{ff})+2f_\nu\otimes P^R_{ff}
+2f_-\otimes P^R_{gf}+f_3\otimes P^R_{gf}\\
-\frac{2\pi}{\alpha_W}\frac{\de f_{+}}{\de t}=
\frac{f_{\bar{e}}+f_{\nu}}{2}\otimes P^R_{fg}
+f_3\otimes P^R_{gg}
+f_+\otimes (P^R_{gg}+2P^V_{gg})
\\
-\frac{2\pi}{\alpha_W}\frac{\de f_{-}}{\de t}=
\frac{f_{\bar{\nu}}+f_{e}}{2}\otimes P^R_{fg}
+f_3\otimes P^R_{gg}
+f_-\otimes (P^R_{gg}+2P^V_{gg})
\\
-\frac{2\pi}{\alpha_W}\frac{\de f_{3}}{\de t}=
\frac{f_{\bar{e}}+f_{e}+f_{\bar{\nu}}+f_{\nu}}{4}\otimes P^R_{fg}
+(f_-+f_+)\otimes P^R_{gg}
+2f_3\otimes P^V_{gg}
\end{cases}
\ea

Even if  two systems (\ref{system},\ref{system1}) are  equivalent, 
 eqs. (\ref{system1})   are  true 5$\times$5 system of differential equations,
 while the reduction to the block diagonal form of (\ref{system}) generates  
 a set of 2$\times$2, 2$\times$2 and 1$\times$1 coupled equations.

To appreciate the block diagonalization,
 let us take   the case of an  initial fermionic parton (an electron).
In this case we have only the $T=0$ and $T=1$ components 
(the fermionic system cannot couple with the $T=2$ projection) so, 
while the use of eqs. (\ref{system}) requires the solution of only the  2$\times$2 plus 2$\times$2 system,
 the use of eqs. (\ref{system1}) always forces to solve the full 5$\times$5 set of eqs.
 This simplification is particularly important when the full particle spectrum of the SM 
is taken into account (see next chapter).

Finally, the last step in obtaining the all order resummed overlap matrix,
requires the evolution of the
$\ff{i}{(T)}$'s according to eqn. (\ref{system}) with appropriate initial
conditions, and inserting the evolved 
$\ff{i}{(T)}$'s into (\ref{fact}). This can by done by exploiting the 
recovered isospin symmetry, which allows us to write: 
\be
\mathop{{\vspace{3pt}\cal O}}\limits^{ i}_{j}
(p_1,p_2;k_1,k_2)=\sum_{T}
\int \frac{dz_1}{z_1}
\frac{dz_2}{z_2}
\sum_{k,l}^{L,R,g}\;
\mathop{{\vspace{3pt} f}}\limits^{k}_{j}\!_{\!\!(T)}(z_1;s,M^2)
\;\mathop{{\vspace{3pt}\cal O}}\limits^{l}_{k}\!^H_{(T)}
(z_1 p_1,z_2 p_2;k_1,k_2)\;
\mathop{{\vspace{3pt} f}}\limits^{i}_{l}\!_{\!\!(T)}(z_2;s,M^2)
\ee

\section{\bf Full EW evolution equations }

Proceeding in analogy with previous section, we now introduce longitudinal
gauge bosons and consider the full SU(2)$\otimes$U(1) electroweak
group. According to the equivalence theorem, we replace longitudinal
gauge bosons with the corresponding Goldstone bosons. 
We choose to work in an axial gauge so that this substitution can be done
without higher order corrections  in the definition of the asymptotic states
\cite{Beenakker:2001kf}.

Our notation goes as follows:
\begin{itemize}
\item
$\F{}{L_i}\!\!_{\alpha\beta}$ represent structure functions 
for left fermions of the $i$-family, including
leptons and quarks
\footnote{Quarks 
structure functions (valid for $L$, $R$ or $LR$ type) are averaged over initial color
$ \F{}{}\equiv\frac{1}{N_c}\sum_{color}\F{}{quarks^{}}\;$
with $N_c=3$.
}
; indices $\alpha,\beta=1,2$ correspond to $\nu,e$ or $u,d$ 
for the first family ($i=1$) and so on. 
$\F{}{\bar{L}_i}\!\!_{\alpha\beta}$ indicates the structure function for the corresponding antifermions.
\item
$\F{}{R_i}$ stand for right fermions  in the $i$th family.
We work in the SM and do not consider right neutrinos.
$\F{}{\bar{R}_i}$ is for right antifermions.
\item
$\F{}{LR_i}$ represent the ``mixed legs'' case where the
  left leg is for a left fermion and the 
right leg for a right fermion of the same charge for the $i$-family, or
  viceversa. 
Such  structure functions are
relevant only for the case of initial transversely polarized beams \cite{transverse}.  
\item
$\F{}{\f}\!_{ab}$ represent structure functions for
the Goldstone ($\f_1,\f_2,\f_3$) - Higgs ($h$) sector. 
 The Goldstone  modes are related to the corresponding longitudinal gauge bosons.
Here $a,b=1,2,3,4$ stand for  $(\f_1,\f_2,\f_3,h)$.
\item
$\F{}{g}\!_{AB}$ stand for transverse $W_A$ gauge bosons belonging to the
  SU(2) sector: $A,B=1,2,3$.

\item ${\cal F}_{\cal B}$ is the structure function for the U(1) ${\cal B}$ gauge boson.

\item
$\F{}{{\cal B}3}$ is the ``mixed leg'' case involving ${\cal B},W_3$ transverse
  gauge bosons.
\end{itemize}

We choose to work in gauge eigenstate basis 
in order to represent the various structure functions.
This basis is  convenient for calculations since at
very high energy we can consider massless propagators and therefore avoid
the complications arising from mass insertions.
Notice however that the asymptotic states appearing in the  $S$ matrix
are truly mass eigenstates, so we need to rotate to the physical base as a
final step.

The leading one loop graphs  contributing, in the axial gauge,  to
 the  splitting functions are shown in  fig.(\ref{figg}).
 The kinematical structure is the same as in  QCD; however
 the group  structure is much more complicated
due  to the absence of isospin averaging on the initial states,
 in contrast with the corresponding
(unbroken) QCD case where physical quantities are averaged over color. 
This   implies  uncanceled infrared singularities and 
the introduction of a new kind of infrared singular splitting functions \cite{full}.

We define $\alpha_W\equiv\frac{g^2}{4\pi},\,
\alpha_Y\equiv\frac{g'^2}{4\pi}$; the various matrices $t^A,{\cal T}^c,\dots$ 
appearing in eqs. (\ref{fulleqs}) are defined in the appendix.
 Y is the hypercharge operator, which is a diagonal matrix with appropriate
eigenvalues in the different representations (left/right fermions and antifermions,
longitudinal gauge bosons and so on). We now write down
the infrared evolution equations  for the structure
functions in matrix form. Family indices here are always understood;
note that the Yukawa coupling contributions proportional to the matrix
${\cal H}$ defined in the appendix are present only for the third family.

\begin{figure}
      \centering
      \includegraphics[height=90mm]
                  {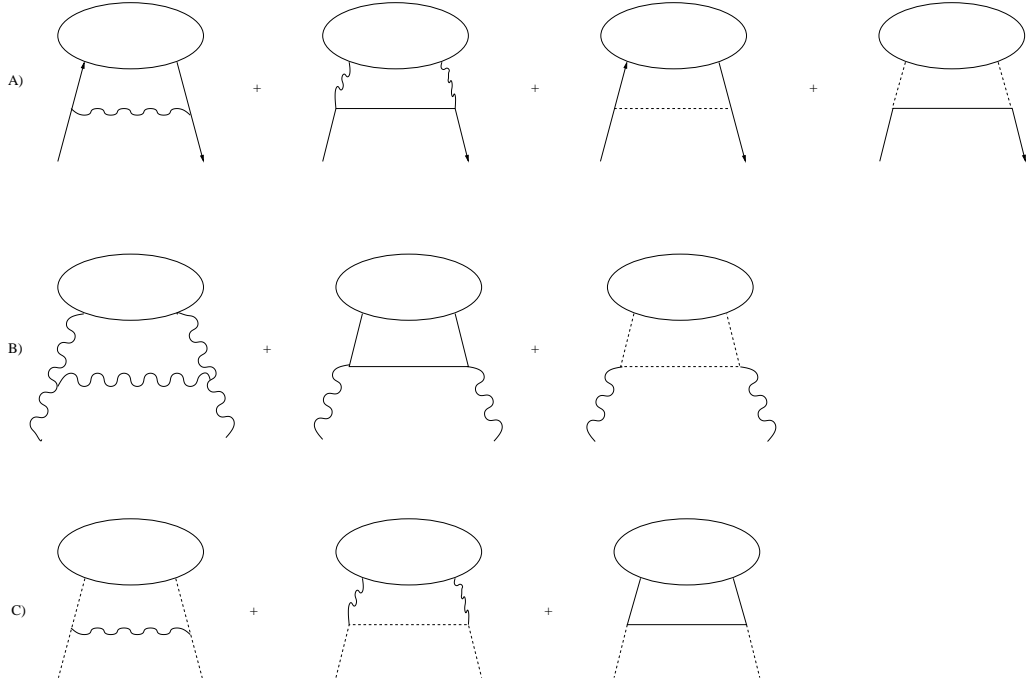}
     
 \caption{\label{figg}
Leading real emission Feynman diagrams in axial gauge:
A) Feynman diagrams contributing to the evolution of the fermionic
structure functions;
B) Feynman diagrams contributing to the evolution of the  transverse
gauge boson structure functions;
C) Feynman diagrams contributing to the evolution of the scalar 
structure functions.
The wavy lines are  transverse gauge bosons, dashed lines stay for Higgs sector particles and 
straight lines for fermions.
 }  
      \end{figure}

\footnotesize\ba\label{fulleqs}
\!\!\!\der\effe{L}\!\!_{\alpha\beta} &=&
\frac{\alpha_W}{2 \pi}\left\{
C_f \effe{L}\!\!_{\alpha\beta}\otimes P^V_{ff}+
(t^{C}  \effe{L}\,^t  t^{C} )_{\beta\alpha}\otimes P^R_{ff}+
 \left(t^B t^A \right)_{\beta \alpha}\;\effe{g}\!\!_{AB}\otimes
P^R_{gf}\right\}+\!\!\!\!\!\!\!\!\!\!\!
\\\nonumber
&&\frac{\alpha_Y}{2 \pi}\,\left\{
\frac{1}{2}
(Y^2\effe{L}+\effe{L}Y^2)_{\alpha\beta}\otimes P^V_{ff}+
 (Y\effe{L}Y)_{\alpha\beta}\otimes
 P^R_{ff}+(Y^2)_{\alpha\beta}\,
 \effe{{\cal B}}\otimes P^R_{gf}\right\}+
\frac{\sqrt{\alpha_W\alpha_Y}}{2\pi}\left\{
(t^3Y+Yt^3)_{\alpha\beta}\,\F{}{{\cal B}3}\otimes P^R_{gf}
\right\}+\\\nonumber
&&
\frac{1}{32\pi^2}\!\left\{
\sum_a({\Psi}^a \cdot{\cal H}\cdot{\cal H}^+  \cdot{\Psi}^{a+})_{\alpha\alpha}\;\effe{L}\!\!_{\alpha\beta}\otimes {\cal P}^V_{LL}+\!
({\Psi}^{a} \cdot{\cal H}\;\effe{R}\!\!_{}\,^{t}\;{\cal H}^+  \cdot{\Psi}^{a+})_{\beta\alpha}\otimes P^R_{RL}
\;+\;
\!({\Psi}^b \cdot{\cal H} \cdot{\cal H}^+  \cdot{\Psi}^{a+})_{\beta\alpha}\;\effe{\f}\!\!_{ab}\otimes P^R_{\f L}
\!\right\}
;\\
 \nonumber
\der\effe{\bar L}\!\!_{\alpha\beta} &=&
\frac{\alpha_W}{2 \pi}\left\{
C_f \effe{\bar L}\!\!_{\alpha\beta}\otimes P^V_{ff}+
(t^{C}  \effe{\bar L}  t^{C} )_{\alpha \beta}\otimes P^R_{ff}+
 \effe{g}\!\!_{AB}\left(t^A t^B \right)_{\alpha\beta}\otimes
P^R_{gf}\right\}+\\\nonumber
&&\frac{\alpha_Y}{2 \pi}\,\left\{
\frac{1}{2}
(Y^2\effe{\bar L}+\effe{\bar L}Y^2)_{\alpha\beta}\otimes P^V_{ff}+
 (Y\effe{\bar L}Y)_{\alpha\beta}\otimes
 P^R_{ff}+(Y^2)_{\alpha\beta}\,
 \effe{{\cal B}}\otimes P^R_{gf}\right\}+
\frac{\sqrt{\alpha_W\alpha_Y}}{2\pi}\left\{
(t^3Y+Yt^3)_{\alpha\beta}\,\F{}{{\cal B}3}\otimes P^R_{gf}
\right\}+\\\nonumber
&&\!
\frac{1}{32\pi^2}\!\left\{
\sum_a({\Psi}^a \cdot{\cal H}\cdot{\cal H}^+  \cdot{\Psi}^{a+})_{\alpha\alpha}\;\effe{\bar L}\!\!_{\alpha\beta}\otimes {\cal P}^V_{LL}+
\!({\Psi}^{a} \cdot{\cal H}\;\effe{\bar R}\!\!_{}\;{\cal H}^+  \cdot{\Psi}^{a+})_{\alpha \beta}\otimes P^R_{RL}+
\!({\Psi}^a \cdot{\cal H} \cdot{\cal H}^+  \cdot{\Psi}^{b+})_{\alpha\beta}\;\effe{\f}\!\!_{ab}\otimes P^R_{\f L}
\right\}
;\\ \nonumber
\der \effe{R}\!\!_{\alpha\beta}&=&
\frac{\alpha_Y}{2 \pi}\,y_{R}^2\left\{  \effe{R} \otimes P^V_{ff}+
 \effe{R} \otimes P^R_{ff}+ \effe{{\cal B}}\otimes P^R_{gf}\right\}+\\\nonumber
&&\!
\frac{1}{32\pi^2}\!\left\{
 \sum_a( {\cal H}^+\cdot {\Psi}^{a+}\cdot{\Psi}^{a}\cdot{\cal H}  )_{\alpha\alpha}\;\effe{R}\!\!_{\alpha\beta}\otimes {\cal P}^V_{RR}+
\!({\cal H}^+\cdot {\Psi}^{a+} \;\effe{L}\!\!_{}^t\;{\Psi}^{a}\cdot{\cal H}  )_{\beta\alpha}\otimes P^R_{LR}+
\!({\cal H}^+\cdot {\Psi}^{b+} \cdot {\Psi}^{a}\cdot{\cal H})_{\beta\alpha} \;\effe{\f}\!\!_{ab} \otimes P^R_{\f R}
\right\};
\\ \nonumber
\der \effe{\bar R}\!\!_{\alpha\beta}&=&
\frac{\alpha_Y}{2 \pi}  \,y_{R}^2 \left\{\effe{\bar R} \otimes P^V_{ff}+
 \effe{\bar R} \otimes P^R_{ff}+ \effe{{\cal B}}\otimes P^R_{gf}\right\}+\\\nonumber
&&\!
\frac{1}{32\pi^2}\!\left\{
 \sum_a( {\cal H}^+\cdot {\Psi}^{a+}\cdot{\Psi}^{a}\cdot{\cal H}  )_{\alpha\alpha}\;\effe{\bar R}\!\!_{\alpha\beta}
\otimes {\cal P}^V_{RR}+\!
({\cal H}^+\cdot {\Psi}^{a+} \;\effe{\bar L}\!\!_{}\;{\Psi}^{a}\cdot{\cal H}  )_{\alpha\beta}\otimes P^R_{LR}+
\!({\cal H}^+\cdot {\Psi}^{a+} \cdot {\Psi}^{b}\cdot{\cal H})_{\alpha\beta} \;\effe{\f}\!\!_{ab} \otimes P^R_{\f R}
\right\};
\\ \nonumber
\der \effe{\f}\!\!_{ab}&=&\frac{\alpha_W}{2 \pi}
\left\{
C_f \effe{\f}\!\!_{ab}\otimes P^V_{\f\f}+
({\cal T}_L^C  \effe{\f}  {\cal T}_L^C )_{a b}\otimes P^R_{\f\f}+
 \effe{g}\!\!_{AB}\left({\cal T}_L^B {\cal T}_L^A \right)_{ b a }\otimes
P^R_{g\f}\right\}+\\\nonumber
&&\frac{\alpha_Y}{2 \pi}\left\{
\frac{1}{2}(Y^2  \effe{\f}+ \effe{\f}Y^2)_{a b}\otimes P^V_{\f\f}+
 (Y\effe{\f}Y)_{a b } 
\otimes P^R_{\f\f} +(Y^2)_{ab} \effe{{\cal B}} \otimes P^R_{g\phi}
\right\}+
\frac{\sqrt{\alpha_W\alpha_Y}}{2\pi}\left\{
({\cal T}^3_LY+Y{\cal T}^3_L)_{ab}\,\effe{{\cal B}3}
\otimes P^R_{g\f}
\right\}+\\\nonumber
&&
\frac{1}{32\pi^2}\left\{
Tr({\Psi}^b \cdot{\cal H} \cdot{\cal H}^+  \cdot{\Psi}^{a+}+
{\Psi}^a \cdot{\cal H} \cdot{\cal H}^+  \cdot{\Psi}^{b+}) 
\effe{\f}\!\!_{ab} \otimes {\cal P}^V_{\f\f}+
\right.\\ \nonumber&&\left.
Tr[({\Psi}^a \cdot {\cal H} \cdot {\cal H}^+  \cdot {\Psi}^{b+})\;\effe{L}\,^t]\otimes P^R_{L \f}+
Tr[({\Psi}^b \cdot{\cal H} \cdot{\cal H}^+  \cdot{\Psi}^{a+})\;\effe{\bar L}]\otimes P^R_{L \f}+
\right.
\\\nonumber
&&\left.
Tr[({\cal H}^+\cdot {\Psi}^{a+} \cdot {\Psi}^{b}\cdot{\cal H}) \;\effe{R}\,^t ] \otimes P^R_{R\f}+
Tr[({\cal H}^+\cdot {\Psi}^{b+} \cdot {\Psi}^{a}\cdot{\cal H}) \;\effe{\bar R}] \otimes P^R_{R\f}
\right\};\\ \nonumber
\der \effe{g}\!\!_{AB}&=&\frac{\alpha_W}{2 \pi}\!
\left\{
C_g \effe{g}\!\!_{AB}\otimes P^V_{gg}+\!
(T_V^C  \effe{g}  T_V^C)_{AB}\otimes P^R_{gg}\!+\!\!
\left(\sum_{L} {\rm Tr}\!\left[t^B \effe{L}\,^t t^A \right]\!\!+\!\!
\sum_{\bar L}  {\rm Tr}\!\left[t^A \effe{\bar L} t^B \right] \right)\!\otimes P^R_{fg}+\!
{\rm Tr}\!\left[{\cal T}_L^B \effe{\f}\,^t 
{\cal T}_L^A \right]\!\otimes P^R_{\f g}\right\};\\\nonumber
\der \effe{{\cal B}}&=&
\frac{\alpha_Y}{2 \pi}
\left\{  
\effe{{\cal B}}\otimes P^R_{{\cal B}{\cal B}}+
\sum_{F=L,\bar{L},R,\bar R} {\rm Tr}\left[Y \effe{F} Y \right]\otimes
P^R_{fg}+ {\rm Tr}\left[Y  \effe{\f}  Y\right]\otimes
P^R_{\f g}\right\};\\\nonumber
\der \effe{{\cal B}3}&=&\frac{\alpha_W}{2 \pi}
\left\{\frac{C_g}{2}\effe{{\cal B}3}\otimes P^V_{gg}\right\}
+\frac{\sqrt{\alpha_W\alpha_Y}}{2 \pi}
\left\{
\sum_{F=L,\bar L} {\rm Tr}\left[Y \effe{F} t^3 \right]\otimes
P^R_{fg}+ {\rm Tr}\left[Y \effe{\f} 
{\cal T}_L^3 \right]\otimes
P^R_{\f g}\right\}+
\frac{\alpha_Y}{2 \pi}
\left\{ \frac{1}{2 }
\;\effe{{\cal B}3}\otimes P^V_{{\cal B}{\cal B}}
\right\}
;\\\nonumber
\der \effe{LR}\!\!_{\alpha\beta}\!&=&\!
\frac{\alpha_W}{2 \pi}\left\{
\frac{C_f}{2} \effe{LR}\!\!_{\alpha\beta}\otimes P^V_{ff} \right\}\!+\!
\frac{\alpha_Y}{2 \pi}\left\{
 \frac{1}{2}(y_{L}^2+y_{R}^2 )\effe{LR}\!\!_{\alpha\beta}
\otimes P^V_{ff}+
y_{L}y_{R} \effe{LR}\!\!_{\alpha\beta}
\otimes P^R_{ff} \right\}\!+\!
\frac{1}{32\pi^2}\!\left\{\frac{1}{2}
\sum_a({\Psi}^a \cdot{\cal H}\cdot{\cal H}^+  \cdot{\Psi}^{a+})_{\alpha\alpha}
\right.\\\nonumber &&\left.
\;\effe{L R}\!\!_{\alpha\beta}\otimes {\cal P}^V_{LL}+
\frac{1}{2}\sum_a( {\cal H}^+\cdot {\Psi}^{a+}\cdot{\Psi}^{a}\cdot{\cal H}  )_{\alpha\alpha}\;\effe{LR}\!\!_{\alpha\beta}
\otimes {\cal P}^V_{RR}+
\sum_a({\cal H}^+\cdot {\Psi}^{a+}\cdot \effe{LR}\,^t
 \cdot{\cal H}^\cro\cdot  {\Psi}^{a+})_{\beta\alpha}
\otimes {\cal P}^R_{LR}\right\}
;\\ \nonumber
\der \effe{\bar{L}\bar{R}}\!\!_{\alpha\beta}&=&
\frac{\alpha_W}{2 \pi}\left\{
\frac{C_f}{2} \effe{\bar{L}\bar{R}}\!\!_{\alpha\beta}\otimes P^V_{ff} \right\}+
\frac{\alpha_Y}{2 \pi}\left\{
 \frac{1}{2}(y_{L}^2+y_{R}^2) \effe{\bar{L}\bar{R}}\!\!_{\alpha\beta}
\otimes P^V_{ff}+
y_{L}y_{R} \effe{\bar{L}\bar{R}}\!\!_{\alpha\beta}
\otimes P^R_{ff} \right\}+
\frac{1}{32\pi^2}\left\{\frac{1}{2}
\sum_a({\Psi}^a \cdot{\cal H}\cdot{\cal H}^+  \cdot{\Psi}^{a+})_{\alpha\alpha}\;
\right.\\\nonumber &&\left.
\effe{\bar L \bar R}\!\!_{\alpha\beta}
\otimes {\cal P}^V_{LL}+
\frac{1}{2}
\sum_a( {\cal H}^+\cdot {\Psi}^{a+}\cdot{\Psi}^{a}\cdot{\cal H}  )_{\alpha\alpha}\;\effe{\bar L \bar R}\!\!_{\alpha\beta}
\otimes {\cal P}^V_{RR}+
\sum_a({\Psi}^{a} \cdot{\cal H}\;\effe{\bar{L}\bar{R}}\cdot
\Psi^{a} \cdot{\cal H})_{\alpha\beta}\otimes P^R_{LR}
\right\}
;\\ \nonumber
\ea\normalsize

We now proceed and reduce eqs. (\ref{fulleqs}) to scalar equations. We work in the
limit of light Higgs; 
the relevant symmetry is then the (recovered) SU(2)$\otimes$ U(1) gauge group, and the
states can be classified according to the total t-channel isospin
${\bf T}$ and the total t-channel hypercharge ${\bf Y}$; however we have to
add the conserved quantum number $CP$ in order to provide a complete 
classification of the states (see Appendix).
 Then, our structure functions are labeled by $({\bf T},{\bf Y},\,CP) $.
Furthermore, we have to consider the ``mixed cases'':
 left fermion-right fermion when we are in presence of transverse 
polarized initial beams \cite{transverse},
 $B-W_3$ mixing when the asymptotic states are $\gamma$'s or transverse $Z$ \cite{full}
and $h-\phi_3$  for longitudinal Goldstone modes \cite{lon}.
While we refer to the cited works for details, here we stress the fact that
we work in the $(\f_1,\f_2,\f_3,h)$ rather than the $1,2,\bar{1},\bar{2}$
basis used in \cite{lon}.
The sum $\sum_i N_c^i$ is over all the fermions where $ N_c^i=1$ for leptons and
$ N_c^i=3$ for quarks.

The scalar evolution equations, using definitions (\ref{inizio}-\ref{fine}) 
and eqs. (\ref{fulleqs}) can finally be written as:

\footnotesize\ba \label{eq1}
\begin{cases}
\der
\ff{R_{i}}{(00;+)}
=
\frac{\alpha_Y}{2 \pi}\;\left(
y_{R}^2\; \ff{R_i}{(0,0;+)}  \otimes \left( P^V_{ff}+P^R_{ff} \right)+
 y_{R}^2\, \ff{\cal B}{(0,0;+)} \otimes P^R_{gf}\right)+
\\\quad\quad\quad\quad\quad\quad
\frac{\delta_{i3}}{32 \pi^2}\; h_R^2\left(
4 \; \ff{R_3}{(0,0;+)}  \otimes {\cal P}^V_{RR}+
4 \; \ff{L_3}{(0,0;+)}  \otimes {P}^R_{LR}+
4 \; \ff{\f}{(0,0;+)}  \otimes {P}^R_{\f R}
\right);
\\
\der \ff{\cal B}{(0,0;+)}=\frac{\alpha_Y}{2 \pi}
\;\left( \ff{\cal B}{(0,0;+)} \otimes P^V_{\cal BB}+
 2 \sum_i\,N_c^i\,\left( 2\,y_{L}^2 \ff{L_i}{(0,0;+)}+
 \sum_R\, y_{R}^2\,\ff{R_i}{(0,0;+)}\right)\otimes P^R_{fg}+
\ff{\f}{(0,0;+)}\otimes P^R_{\phi g}\right);
\\
\der \ff{L_{i}}{(0,0;+)}=
\frac{\alpha_W}{2 \pi}
\;\left(
\frac{3}{4}\; \ff{L_{i}}{(0,0;+)}\otimes \left( P^V_{ff}+P^R_{ff} \right)+
 \frac{3}{4}\, \ff{g}{(0,0;+)}\otimes P^R_{gf}\right)+
\frac{\alpha_Y}{2 \pi}
\;\left(
y_{L}^2\; \ff{L_{i}}{(0,0;+)}\otimes \left( P^V_{ff}+P^R_{ff} \right)+
\, y_{L}^2\, \ff{\cal B}{(0,0;+)}\otimes P^R_{gf}\right)+
\\\quad\quad\quad\quad\quad\quad
\frac{\delta_{i3}}{32 \pi^2}\;\left(
2\,(h_t^2+h_b^2)\; \ff{L_{3}}{(0,0;+)}  \otimes {\cal P}^V_{LL}+
2\,\sum_R\,h_R^2\, \ff{R_3}{(0,0;+)}
 \otimes {P}^R_{RL}+
2\,(h_t^2+h_b^2)\; \ff{\f}{(0,0;+)}  \otimes { P}^R_{\f L}
\right);
\\
\der \ff{g}{(0,0;+)}=\frac{\alpha_W}{2 \pi}
\;\left(     2\;\ff{g}{(0,0;+)}\otimes (P^V_{gg}+P^R_{gg})+
\sum_i \,N_c^i\,\ff{L_i}{(0,0;+)}\otimes P^R_{fg}+
 \ff{\f}{(0,0;+)}\otimes P^R_{\f g}\right);\\
\der \ff{\f}{(0,0;+)}=\frac{\alpha_W}{2 \pi} 
\;\left( \frac{3}{4}\, \ff{\phi}{(0,0;+)}\otimes \left(P^V_{\phi\phi} +P^R_{\phi\phi}\right)
 + \frac{3}{4}\,\ff{g}{(0,0;+)}\otimes P^R_{g\phi} \right)+
\frac{\alpha_Y}{2 \pi} 
\;\left(  \frac{1}{4}  \, \ff{\phi}{(0,0;+)}\otimes 
(P^R_{\phi\phi}+ P^V_{\phi\phi}  ) 
+ \frac{1}{4}\,\ff{\cal B}{(0,0;+)}\otimes P^R_{g\phi} \right)+
\\\quad\quad\quad\quad\quad\quad
\frac{N_c}{32 \pi^2}\;\left(
2\,(h_t^2+h_b^2)\; \ff{\f}{(0,0;+)}  \otimes {\cal P}^V_{\f \f}+
2(h_t^2+h_b^2)\; \ff{L_3}{(0,0;+)}  \otimes { P}^R_{L \f}+
2\,\,\sum_R\,h_R^2\, \ff{R_3}{(0,0;+)}
 \otimes {P}^R_{R\f}
\right);
\end{cases}
\ea

\ba
\begin{cases}
\label{scalar1}
\der \ff{R_i}{(0,0;-)}=
\frac{\alpha_Y}{2 \pi}\;\left(
y_{R}^2\; \ff{R_i}{(0,0;-)}\otimes \left( P^V_{ff}+P^R_{ff} \right)\right)
+\\
\quad\quad\quad\quad\quad\quad
\frac{\delta_{i3}}{32 \pi^2}\,h_R^2\;\left(
4\, \ff{R_3}{(0,0;-)}  \otimes {\cal P}^V_{RR}+
4\,\ff{L_3}{(0,0;-)}  \otimes { P}^R_{ LR}\pm 4\,\ff{\f}{(0,0;-)}  \otimes { P}^R_{ \f R}
\right) \quad\quad(${\rm + for up and - for down type fermions}$)
;
\\
\der \ff{L_i}{(0,0;-)}=
(\frac{3}{4}\frac{\alpha_W}{2 \pi}+y_{L}^2\frac{\alpha_Y}{2 \pi})
\ff{L_i}{(0,0;-)}\otimes(P_{ff}^V +P^R_{ff})+\\
\quad\quad\quad\quad\quad\quad
\frac{\delta_{i3}}{32 \pi^2}\;\left(
2\,(h_t^2+h_b^2)\, \ff{L_3}{(0,0;-)}  \otimes {\cal P}^V_{LL}+
2\,\sum_R\,h_R^2\, \ff{R_3}{(0,0;-)}
 \otimes { P}^R_{ RL}-2\,(h_t^2-h_b^2)\,\ff{\f}{(0,0;-)}  \otimes { P}^R_{ \f L}
\right) ;
\\
\der \ff{\phi}{(00;-)}=\frac{\alpha_W}{2 \pi} 
\;\left(
 \frac{3}{4}\, \ff{\phi}{(00;-)}\otimes \left(P^V_{\phi\phi} +P^R_{\phi\phi}\right)\right)+
\frac{\alpha_Y}{2 \pi} 
\;\left(\frac{1}{4}  \,
  \ff{\phi}{(00;-)}\otimes 
\left(P^V_{\phi\phi} +P^R_{\phi\phi}\right)
\right) +\\
\quad\quad\quad\quad\quad\quad
\frac{N_c}{32 \pi^2}\;\left(
2\,(h_t^2+h_b^2)\, \ff{\f}{(0,0;-)}  \otimes {\cal P}^V_{\f\f}-2
(\,h_b^2\,\ff{R_{d_3}}{(0,0;-)} -\,h_t^2\,\ff{R_{u_3}}{(0,0;-)})  \otimes { P}^R_{ R\f}-2
\,(h_t^2-h_b^2)\,\ff{L_3}{(0,0;-)}  \otimes { P}^R_{ L\f}
\right)   ; 
  \end{cases}                     
\ea

\ba
\begin{cases}
\der \ff{L_i}{(1,0;+)}=\frac{\alpha_W}{2 \pi} 
\;\left(  \ff{L_i}{(1,0;+)}\otimes P_{ff}^V -\frac{1}{4} \ff{L_i}{(1,0;+)}\otimes  
\left( P^V_{ff}+P^R_{ff} \right) \right)+
\frac{\alpha_Y}{2 \pi}\;\left(
y_{L}^2\; \ff{L_i}{(1,0;+)}\otimes \left( P^V_{ff}+P^R_{ff} \right)
 \right) +\\\quad\quad\quad\quad\quad\quad
  \frac{\sqrt{\alpha_W\alpha_Y}}{2\pi}\;\left(
y_{L}\;\ff{B3}{(1,0;+)} \otimes  P^R_{gf}
\right) +
\frac{\delta_{i3}}{32 \pi^2}\;\left(
2\,(h_t^2+h_b^2)\; \ff{L_3}{(1,0;+)}  \otimes {\cal P}^V_{LL}-
2\,(h_t^2-h_b^2)\; \ff{\f}{(1,0;+)}  \otimes { P}^R_{ \f L}
\right)  ;
\\
\der \ff{B3}{(1,0;+)}= \frac{\alpha_W}{2 \pi}
\;\left( 
\frac{1}{2}\ff{B3}{(1,0;+)}\otimes P^V_{gg}\right)+
\frac{\sqrt{\alpha_W\alpha_Y}}{2\pi}
\;\left(  4\,y_{L}\sum_i\,N_c^i\, \ff{L_i}{(1,0;+)}\otimes P^R_{fg}+
   \ff{\phi}{(1,0;+)}\otimes P^R_{\phi g}\right)+
\frac{\alpha_Y}{2 \pi}
\,\left( \frac{1}{2}\ff{B3}{(1,0;+)} \otimes P^V_{BB}\right)
;
\\
\der \ff{\phi}{(10;+)}=
\frac{\alpha_W}{2 \pi} 
\;\left( \ff{\phi}{(10;+)}\otimes P^V_{\phi\phi}- \frac{1}{4}\, \ff{\phi}{(10;+)}\otimes 
\left(P^V_{\phi\phi} +P^R_{\phi\phi}\right)
\right)+
\frac{\alpha_Y}{2 \pi} 
\;\left(\frac{1}{4}  \,
  \ff{\phi}{(1,0;+)}\otimes 
\left(P^V_{\phi\phi} +P^R_{\phi\phi}\right)
\right)+\\\quad\quad\quad\quad\quad\quad
\frac{\sqrt{\alpha_W\alpha_Y}}{2\pi} \frac{1}{2}\ff{B3}{(1,0;+)}\otimes
P^R_{g\phi} +
\frac{N_c}{32 \pi^2}\;\left(
2\, (h_t^2+h_b^2)\,\ff{\f}{(1,0;+)}  \otimes {\cal P}^V_{\f \f}+(h_b^2-h_t^2)\, \ff{L_3}{(1,0;+)}  \otimes { P}^R_{L \f}
\right)  
;      
\end{cases}
\ea

\ba
\begin{cases}
\der \ff{L_i}{(1,0;-)}=\frac{\alpha_W}{2 \pi}
\;\left(
 \ff{L_i}{(1,0;-)}\otimes P_{ff}^V -
\frac{1}{4} \ff{L_i}{(1,0;-)}\otimes  \left( P^V_{ff}+P^R_{ff} \right) 
+\frac{1}{2}\, \ff{g}{(1,0;-)}\otimes P^R_{gf} \right)  +\\\quad\quad\quad\quad\quad\quad
\frac{\alpha_Y}{2 \pi}\;\left(
y_{L}^2\; \ff{L_i}{(1,0;-)}\otimes \left( P^V_{ff}+P^R_{ff} \right)\right)+
\frac{\delta_{i3}}{32 \pi^2}\,(h_t^2+h_b^2)\;\left(2\, \ff{L_3}{(1,0;-)}  \otimes {\cal P}^V_{LL}+
2\, \ff{\f}{(1,0;-)}  \otimes { P}^R_{\f L}
\right);
\\
\der \ff{g}{(1,0;-)}=\frac{\alpha_W}{2 \pi} 
\;\left( 
    \ff{g}{(1,0;-)}\otimes
 P_{gg}^V + \ff{g}{(1,0;-)}\otimes \left(P^V_{gg}+ P^R_{gg} 
\right)+\sum_i\,N_c^i\,\ff{L_i}{(1,0;-)}\otimes  P^R_{fg}+
 \ff{\phi}{(1,0;-)}\otimes P^R_{\phi g}\right);\\
\der \ff{\phi}{(10;-)}=
\frac{\alpha_W}{2 \pi} \left(
\ff{\phi}{(10;-)}\otimes P^V_{\phi\phi}
- \frac{1}{4}\,  \ff{\phi}{(1,0;-)}\otimes 
\left(P^V_{\phi\phi} +P^R_{\phi\phi}\right)
 + \frac{1}{2}\,\ff{g}{(1,0;-1)}\otimes P^R_{g\phi} \right)  +\\\quad\quad\quad\quad\quad\quad
\frac{\alpha_Y}{2 \pi} 
\;\left(\frac{1}{4}  \,
 \, \ff{\phi}{(10;-)}\otimes 
\left(P^V_{\phi\phi} +P^R_{\phi\phi}\right)
\right)   +
\frac{N_c}{32 \pi^2}\;\left(
2\,(h_t^2+h_b^2)\, \ff{\f}{(1,0;-)}  \otimes {\cal P}^V_{\f \f}+2
(h_t^2+h_b^2)\, \ff{L_3}{(1,0;-)}  \otimes { P}^R_{L \f}
\right)   ;     
\end{cases}
\ea

\ba
\der \ff{g}{(2,0;+)}&=&\frac{\alpha_W}{2 \pi} 
\;\left( 3\,\ff{g}{(2,0;+)}\otimes P_{gg}^V -  \ff{g}{(2,0;+)}\otimes (P^V_{gg}+P^R_{gg}) \right) ;
 \\
\der \ff{\phi}{(11;-)}&=&
\frac{\alpha_W}{2 \pi} 
\;\left( \ff{\phi}{(11;-)}\otimes P^V_{\phi\phi}- \frac{1}{4}\, \ff{\phi}{(11;-)}\otimes 
\left(P^V_{\phi\phi} +P^R_{\phi\phi}\right)
\right)+\\\nonumber
&&\frac{\alpha_Y}{2 \pi} 
\;\left(
 \frac{1}{2}  \,\, \ff{\phi}{(11;-)}\otimes P^V_{\phi\phi}-\frac{1}{4}  \, \ff{\phi}{(11;-)}
\left(P^V_{\phi\phi} +P^R_{\phi\phi}\right)
\right)  +
\frac{N_c}{32 \pi^2}\;\left(
2\,(h_t^2+h_b^2)\, \ff{\f}{(1,1;-)}  \otimes {\cal P}^V_{\f\f}\right) ;           
\\\label{endeq}
\der \ff{\phi}{(11;+)}&=&
\frac{\alpha_W}{2 \pi} 
\;\left( \ff{\phi}{(11;+)}\otimes P^V_{\phi\phi}- \frac{1}{4}\, \ff{\phi}{(11;+)}\otimes 
\left(P^V_{\phi\phi} +P^R_{\phi\phi}\right) \label{scalar2}
\right)+\\\nonumber     
&&\frac{\alpha_Y}{2 \pi} 
\;\left(
 \frac{1}{2}  \,\,  \ff{\phi}{(11;+)}\otimes P^V_{\phi\phi}-\frac{1}{4}  \, \ff{\phi}{(11;+)}
\left(P^V_{\phi\phi} +P^R_{\phi\phi}\right)
\right)+
\frac{N_c}{32 \pi^2}\;\left(
2\,(h_t^2+h_b^2)\, \ff{\f}{(1,1;+)}  \otimes {\cal P}^V_{\f\f}\right) ;\label{eqn}
\ea

\ba
\begin{cases}
\der \ff{LR_i}{\!\!_u}{(\frac{1}{2},\frac{1}{4} ;\pm)}=\frac{\alpha_W}{2 \pi}
\;\left(  \frac{3}{8}\, \ff{LR_i}{\!\!_u}{(\frac{1}{2},\frac{1}{4};\pm)}\otimes P_{ff}^V \right)    +
\frac{\alpha_Y}{2 \pi}
\;\left(\frac{1}{2}(y_L-y_{R}^u)^2\; \ff{LR_i }{\!\!_u}{(\frac{1}{2},\frac{1}{4} ;\pm)}\otimes P^V_{ff}+
y_L\,y_{R}^u \; \ff{LR_i }{\!\!_u}{(\frac{1}{2},\frac{1}{4} ;\pm)}\otimes (P^V_{ff}+P^R_{ff})
\right)+\\
\quad\quad\quad\quad\quad\quad\quad\quad\quad\frac{\delta_{i3}}{32 \pi^2}\;\left(
2\,h_t^2\;\ff{LR_3}{\!\!_u}{(\frac{1}{2},\frac{1}{4} ;\pm)}\otimes {\cal P}^V_{RR}+
\,(h_t^2+h_b^2)\;\ff{LR_3}{\!\!_u}{(\frac{1}{2},\frac{1}{4} ;\pm)}\otimes {\cal P}^V_{LL}
-2 h_t\,h_b\;\ff{LR_3}{\!\!_d}{(\frac{1}{2},\frac{1}{4} ;\pm)}
\otimes {\cal P}^R_{LR}
\right)
;
\\
\der \ff{LR_i}{\!\!_d}{(\frac{1}{2},\frac{1}{4} ;\pm)}=\frac{\alpha_W}{2 \pi}
\;\left(  \frac{3}{8}\, \ff{LR_i}{\!\!_d}{(\frac{1}{2},\frac{1}{4};\pm)}\otimes P_{ff}^V \right)    +
\frac{\alpha_Y}{2 \pi}
\;\left(\frac{1}{2}(y_{L}-y_R^d)^2\; \ff{LR_i }{\!\!_d}{(\frac{1}{2},\frac{1}{4} ;\pm)}\otimes P^V_{ff}+
y_L\,y_{R}^d \; \ff{LR_i }{\!\!_d}{(\frac{1}{2},\frac{1}{4} ;\pm)}\otimes (P^V_{ff}+P^R_{ff})
\right)+\\
  \quad\quad\quad\quad\quad\quad\quad\quad\quad \frac{\delta_{i3}}{32 \pi^2}\;\left(
2\,h_b^2\;\ff{LR_3}{\!\!_d}{(\frac{1}{2},\frac{1}{4} ;\pm)}\otimes {\cal P}^V_{RR}+
\,(h_t^2+h_b^2)\;\ff{LR_3}{\!\!_d}{(\frac{1}{2},\frac{1}{4} ;\pm)}\otimes {\cal P}^V_{LL}
-2 h_t\,h_b\;\ff{LR_3}{\!\!_u}{(\frac{1}{2},\frac{1}{4} ;\pm)}
\otimes {\cal P}^R_{LR}
\right)
\end{cases}
\ea
\normalsize

By looking at the above equations 
we can see
 that only structure functions that share
 the same set of quantum numbers 
are mixed by the evolution equations.

Perturbative initial conditions are independent of
the particular process considered; in
fact they are  set  at the hard scale $\sqrt{s}$, where no soft radiation
is emitted, corresponding
to tree level values for $\effe{}$. 

We have therefore (see fig.1):
\be\label{ini1}
\F{i}{j}\!_{\alpha'\beta'}^{\alpha\beta}(x,\mu=\sqrt{s})
=\delta_{ij}\delta_{\alpha\alpha'}\delta_{\beta \beta'}\delta(1-x)
\ee
or, expressed in rotated structure functions with total isospin and hypercharge
 \be\label{ini2}
\mathop{{\vspace{3pt}f}}\limits_{i}^{j}\!{\scriptscriptstyle (T,Y;CP)}
(x,\mu=\sqrt{s})
=\delta_{ij}\;\delta(1-x)
\ee
as can be checked by their definition in eqs.(\ref{inizio}-\ref{fine}).

Eqs. (\ref{fulleqs}-\ref{endeq}), together with the initial conditions
(\ref{ini1}-\ref{ini2}) are the main results of this paper.
 They are the electroweak analogous of DGLAP equations for QCD, and allow
 to calculate  the distribution of a particle $i$ inside a
particle $j$ at a given scale $\mu^2 \geq M_W^2$. 
Only electroweak effects are treated here;
 a complete treatment needs to take into account also the low energy ($\mu^2 \leq M_W^2$)
 QED running and the QCD effects of course.

As stated above, we choose to work in a gauge basis since in the high
energy limit mass insertions (but not the top Yukawa  insertions one) can be neglected,
 and calculations are simpler than in the mass eigenstates basis. However, initial (asymptotic) states
are mass eigenstates and we need to rotate to this base in order to obtain
physical cross sections. This can be done as a final rotation to the mass
basis,  once the evolved overlap matrix $\mathop{{\vspace{3pt}\cal O}}\limits^{ i}_{j}
(p_1,p_2;k_1,k_2)$ in the gauge basis is obtained.
 We write here an example in order to clarify this procedure; an  analysis at the 
leading double log level  has already been done
\cite{lon,abelian,full,transverse}. 
We discuss the change of basis
for the structure functions $\F{}{}$, but things work in exactly the same
way for the overlap matrix. 

Let us consider  the case of an initial photon beam
( $| \gamma \ket= c_W|{\cal B} \ket+s_W|W^3 \ket $). The structure function in the
mass eigenstate basis can be written
in terms of structure functions of the corresponding gauge eigenstates:  
\be \label{gamgam}
\F{}{g}\!_{\gamma\gamma}=
s_W^2 \F{}{g}\!_{33}+2 s_W c_W \F{}{{\cal B}3}+c_W^2 \F{}{\cal B}=
s_W^2\ff{g}{(0,0;+)}+c_W^2\ff{\cal B}{(0,0;+)}+
2 s_W c_W \ff{{\cal B}3}{(1,0;+)}-
\frac{4}{3}s_W^2 \ff{g}{(2,0;+)}
\ee
obtained expanding the direct product $(c_W|{\cal B} \ket+s_W|W^3
\ket)\otimes (c_W|{\cal B} \ket+s_W|W^3 \ket)$,
 and where we used (\ref{inizio}-\ref{fine}) to write the $\F{}{}$ in terms of $\ff{}{(T,Y;CP)}$.

\vspace{0.5cm}

{\rm \bf Acknowledgments}: We are grateful to {\bf Marcello Ciafaloni}
 for discussions and  participation to the first part of our  results.

\newpage

\section{Appendix:}

{\bf \Large Kernels of the evolution equations}

\be\label{pff}
P^V_{ff}(z,\kt)=-\delta(1-z)\left(\log\frac{s}{\kt^2}-\frac{3}{2}\right );\;\;\;\; 
P^R_{ff}(z)=\frac{1+z^2}{1-z};\;\;\;\;
\ee 

\ba\label{pgg}
P^V_{\f\f}(z,\kt)=-\delta(1-z)\left(\log\frac{s}{\kt^2}-2 \right);\;\;\;\; 
P^R_{\f\f}(z)=2\frac{z}{1-z};\;\;\;\;
\ea

\ba\label{pww}
P^V_{gg}(z,\kt)&=&-\delta(1-z)\left(\log\frac{s}{\kt^2}-(  \frac{11}{6}-\frac{n_f}{6}- \frac{n_s}{24} ) \right);\;\;\;\; 
P^R_{gg}(z)=2\left(z(1-z)+\frac{z}{1-z}+\frac{1-z}{z}  \right);\\\nonumber
P^V_{{\cal B}{\cal B}}(z)&=&-\left(2\,n_f(2 y_L^2+y_{R_u}^2+y_{R_d}^2)+y_{\f}^2\right)\delta(1-z)\frac{1}{3}
\ea

\ba
P^R_{gf}(z)&=& \frac{1+(1-z)^2}{z},\;\;\;\;
P^R_{g\f}(z)=2 \frac{(1-z)}{z}\;\;\;\;\\
P^R_{fg}(z)&=&z^2+ (1-z)^2;\;\;\;\;
P^R_{\f g}(z)=z (1-z)
\ea

\ba\label{pscalar}
{\cal P}^V_{LL}(z)&=&-\delta(1-z) \;\frac{1}{2}
;\;\;\;\; {\cal P}^V_{RR}(z)=-\delta(1-z)\;\frac{1}{2};\;\;\;\;
{\cal P}^V_{\f\f}(z)=-\delta(1-z)
\\
 P^R_{LR}(z)&=&P^R_{RL}(z)=(1-z);\;\;\;\;\;\;\;\;
P^R_{L\f}(z)=P^R_{R\f}(z)=1;\;\;\;\;
P^R_{\f L}(z)=P^R_{\f R}(z)=z
\nonumber
\ea
where $n_f=\sum_i N_c^i$ is the sum over all the fermion families ($N_c^i=1$ for leptons and $N_c^i=3$ for quarks  ),
$n_s=1$ is the number of Higgs doublets
and
$N_c=3$ is a color factor.

\vspace{1.cm}

{\bf \Large Fermion-Scalar Sector Yukawa Interactions
} 

\vspace{1.cm}

Parametrizing the scalar fields in the following way
\be\label{higgs}
\Phivet=
\frac{1}{\sqrt{2}}(h+i\tau_a\f_a)
=\left(\begin{array}{cc}
\f_0&i\f_+\\i\f_-&\f_0^*\end{array}\right)
=\frac{1}{\sqrt{2}}\left(\begin{array}{cc}
h+i\f_3&\f_2+i\f_1\\-\f_2+i\f_1&h-i\f_3\end{array}\right)
\ee
the interaction between the scalar sector and the Left-Right fermions can be written as
\be
\bar{Q}_L\;\Phivet\; {\cal H}\; Q_R+\bar{Q}_R\; {\cal H}^+ \;\Phivet^+\; Q_L=
\sum_{a=1}^4\;\f^a\left( \bar{Q}_L \;{\Psi}^a \cdot{\cal H}\; Q_R+
\bar{Q}_R\; {\cal H}^+ \cdot{\Psi}^{a+}\;\Phivet^+\; Q_L
\right)
\ee
where ${\Psi}^a=i \tau^a $  for a=1,2,3  and  ${\Psi}^4=I_{2\times 2}$ and
the third family Yukawa couplings are arranged in the
following matricial form
\be
{\cal H}=\left(\begin{array}{cc}
h_t & 0 \\ 0 & h_b\end{array}\right)
\ee

\newpage

{\bf \Large States classification  } 

\vspace{0.5cm}
With the above parameterization (\ref{higgs}),
the Higgs sector of the Standard Model includes the three Goldstone modes 
$\f_1,\f_2,\f_3$ and the Higgs field $h$.
It is  useful to arrange these four states
in the complex form of the doublet/antidoublet matrix
\be\label{high}
\frac{1}{\sqrt{2}}\left(\begin{array}{cc}
h+i\f_3&\f_2+i\f_1\\-\f_2+i\f_1&h-i\f_3\end{array}\right)
=\left(\begin{array}{cc}
\bar{1}&{1}\\\bar{2}&{2}\end{array}\right)
\ee
which transforms as
$
\Phivet\rightarrow \exp[i\alpha_L^At^A]\Phivet
\exp[-i\alpha_R^Bt^B]
$
under the $ SU(2)_L \otimes SU(2)_R$
group. The indices $1,2,\bar{1},\bar{2}$ here refer to the transformation
properties under SU(2)$_L$.

$t^A$ and $T^A, A=1,2,3$ are the $SU(2)_L$ generators in the fundamental and
adjoint representation. 

The other matrices appearing in the text are the
generators of the  $ SU(2)_L \otimes SU(2)_R$ group in the
$(\f_1,\f_2,\f_3,h)$ basis; their explicit form is as follows:
\be
{\cal T}_L^A=\frac{1}{2}\;\left(\begin{array}{cc}
i\, \epsilon_{BAC}& -i\, \delta_{AB}\\
 i\, \delta_{AC}&0\end{array}\right)\qquad
{\cal T}_R^A=\frac{1}{2}\;\left(\begin{array}{cc}
i\, \epsilon_{BAC}& i\, \delta_{AB}\\
- i\, \delta_{AC}&0\end{array}\right)\qquad
\ee
Furthermore, 
$({\cal T}_h^3)_{ab}=i(\delta_{4a}\delta_{3b}-\delta_{3a}\delta_{4b})$ and 
$({\cal T}_y^3)_{ab}=\delta_{3a}\delta_{4b}
+\delta_{3b}\delta_{4a}$.

States in the Higgs sector can be classified according to their  
$SU(2)_L \otimes SU(2)_R$ properties, with definite 
$|{\bf T},T^3_L;{\bf T}_R,T^3_R\ket$ quantum numbers (see also appendix
in \cite{lon}). However not all of
the 16 possible states appear in the evolution equations. In fact, a
physical cross section is always an overlap matrix element with a given 
particle in the left leg and its antiparticle in the right one. By charge
conservation then, all the states involved in the evolution equations have
t-channel charge equal to zero: $Q=T^3_L+Y=T^3_L+T^3_R=0$. Then, only 6
states are selected:

\begin{equation}\label{states}\begin{array}{lll}
|1,-1;1,1\ket=-|22\ket\qquad
&
|1,1;1,-1\ket=-|\bar{1}\bar{1}\ket\qquad
&
|1,0;1,0\ket=-\frac{|1\bar{2}\ket+|\bar{1}2\ket+
|2\bar{1}\ket+|\bar{2}1\ket}{2}
\\
|1,0;0,0\ket=-\frac{|1\bar{2}\ket-|\bar{1}2\ket+
|2\bar{1}\ket-|\bar{2}1\ket}{2}\qquad
&
|0,0;1,0\ket=-\frac{|1\bar{2}\ket+|\bar{1}2\ket-
|2\bar{1}\ket-|\bar{2}1\ket}{2}\qquad
&
|0,0;0,0\ket=-\frac{|1\bar{2}\ket-|\bar{1}2\ket-
|2\bar{1}\ket+|\bar{2}1\ket}{2}
\end{array}\ee

In the high energy limit we are considering the $SU(2)_L\otimes U(1)$
symmetry is recovered, and states evolve according to their
total t-channel isospin ${\bf T}$ and hypercharge $Y$. However
quantum numbers $| {\bf T}, Y\ket$ do not provide a complete
classification of the states. For instance, we have two states
corresponding to $| {\bf T}=0, Y=0\ket$: these are the states 
$|0,0;0,0\ket$ and $|0,0;1,0\ket$
appearing in (\ref{states}). We choose to add another quantum number, CP,
that acts  as\footnote{notice that the definition of CP given in \cite{lon}
  is different from the one used here}:
\be
1\leftrightarrow - \bar{2}\qquad 2\leftrightarrow \bar{1}
\quad\Rightarrow\quad
\phi_1\leftrightarrow -\phi_1\qquad
\phi_2\leftrightarrow \phi_2\qquad
\phi_3\leftrightarrow -\phi_3\qquad
\phi_4\equiv h\leftrightarrow \phi_4\equiv h\qquad
\phi_+
\leftrightarrow -\phi_-
\ee

We can now write the states with given
$|{\bf T},Y^2\ket^{(CP)}$ quantum numbers in terms of 
the states  $|{\bf T},T^3_L;{\bf T}_R,T^3_R\ket$ 
classified according to their $SU(2)_L\otimes SU(2)_R$ 
properties and given in (\ref{states}):
\be
|0,0\ket^{(+)}=|0,0;0,0\ket\qquad
|1,0\ket^{(-)}=|1,0;0,0\ket\qquad
|0,0\ket^{(-)}=|0,0,;1,0\ket
\ee
\be
|1,0\ket^{(+)}=|1,0;1,0\ket\qquad
|1,1\ket^{(+)}=\frac{1}{2}(|1,1;1,-1\ket+|1,-1;1,1\ket)\qquad
|1,1\ket^{(-)}=\frac{1}{2}(|1,1;1,-1\ket-|1,-1;1,1\ket)
\ee

It is now easy, by using (\ref{high}), to write the above states in the
$\phi_1,\phi_2,
\f_3,h$ base and  to find out the corresponding $f$ functions. For instance:
\be
|0,0\ket^{(+)}=|0,0;0,0\ket=-\frac{|1\bar{2}\ket-|\bar{1}2\ket-
|2\bar{1}\ket+|\bar{2}1\ket}{2}=|\f_+\f_-\ket+|\f_-\f_+\ket+
|\f_3\f_3\ket+|\f_h\f_h\ket
\ee
This combination corresponds to
\be
\ff{\f}{\scriptstyle(0,0;+)}=
\frac{\effe{\f}\!\!_{+-}+
\effe{\f}\!\!_{-+}+
\effe{\f}\!\!_{33}+
\effe{\f}\!\!_{hh}}{4}
\ee
In the last step we have written $\ff{\f}\!{\scriptstyle(0,0;+)}$ as a trace over the
lower indices of the ${\cal F}$ operator, in order to simplify  the task of
writing scalar evolution equations.

We use $SU(2)\otimes U(1)$ and CP quantum numbers also to classify states in the
fermions/antifermions and gauge bosons sectors. CP acts as follows:
\be
\nu\leftrightarrow \nu^*\qquad e\leftrightarrow e^*\qquad
W_\mu^+\leftrightarrow -W_\mu^-\qquad W_\mu^3\leftrightarrow -W_\mu^3
\qquad B_\mu\leftrightarrow -B_\mu
\ee
Notice that in the fermionic sector, since CP transforms a fermion in its
own antifermion, the states with defined $SU(2)\otimes U(1)$ properties, as
defined in \cite{lon}:
\be
f_L^{(0)}=\frac{f_e+f_{\nu}}{2}\qquad
f_{\bar L}^{(0)}=
\frac{f_{\bar e}+f_{\bar \nu}}{2}\qquad
f_L^{(1)}=\frac{f_{\nu}-f_e}{2}\qquad
f_{\bar L}^{(1)}=\frac{f_{\bar e}-f_{\bar \nu}}{2}
\ee
do not have definite CP values since CP transforms, for instance,  
$f_L^{(0)}$ into $f_{\bar L}^{(0)}$. 

SU(2) symmetry acts on the gauge bosons sector as a triplet, and
classifying the states does not present any particular difficulty. We can
then proceed and write all the states in the fermionic, scalar, gauge bosons
sectors according to their $|{\bf T},Y^2\ket^{CP}$ quantum numbers. Here we
write directly the corresponding structure functions $f^{(\Ttot,\Ytot;CP)}$ and 
for Left fermions we write
as example the case with $L_i$ equal to left leptons of the first $i=1$ family ($e$ and $\nu$):

\begin{itemize}
\item  singlet states 
$\Ttot=0,\Ytot=0 $
\ba\label{inizio}
\ff{\cal B}{\scriptstyle(0,0;+)}&\equiv& \F{}{\cal B}
\\
\ff{R_i}{\scriptstyle(0,0;+)}&\equiv&  \frac{1}{2}(\effe{R_i}+\effe{\bar{R}_i})
\\
\ff{L_i}{\scriptstyle(0,0;+)}&\equiv&  
 \frac{1}{4}\, {\rm Tr}\left[\effe{L_i}+\effe{\bar{L}_i}\right]
=\frac{1}{4}\left(
\effe{L}\!\!_{e e}+\effe{L}\!\!_{\nu\nu}+\effe{L}\!\!_{\bar e\bar e}
+\effe{L}\!\!_{\bar \nu\bar \nu}\right)
\\
\ff{\f}{\scriptstyle(0,0;+)}&\equiv& \frac{1}{4}\,{\rm Tr}[\F{}{\f}]=
\frac{\effe{\f}\!\!_{+-}+\effe{\f}\!\!_{-+}+\effe{\f}\!\!_{33}+\effe{\f}\!\!_{hh}}{4}
\\
\ff{g}{\scriptstyle(0,0;+)}&\equiv&\frac{1}{3}\, {\rm Tr}[\F{}{g}]=
\frac{\effe{g}\!\!_{+-}+\effe{g}\!\!_{33}+\effe{g}\!\!_{-+}}{3}
\\\nonumber
\\\nonumber
\\
\label{00-1}
\ff{R_i}{\scriptstyle(0,0;-)}&\equiv&  \frac{1}{2}(\effe{R_i}-\effe{\bar{R}_i})
\\
\label{00-2}
\ff{L_i}{\scriptstyle(0,0;-)}&\equiv&   \frac{1}{4}\, {\rm Tr}\left[
\F{}{L_i}-\F{}{\bar{L}_i}\right]
=\frac{1}{4}\left(( \effe{L}\!\!_{ee}+\effe{L}\!\!_{\nu\nu})-(\effe{L}\!\!_{\bar e \bar e}+
\effe{L}\!\!_{\bar \nu \bar \nu})\right)  \\
\label{00-3}
\ff{\f}{\scriptstyle(0,0;-)}&\equiv&-
\frac{1}{2} {\rm Tr}[{\cal T}_R^3\F{}{\f}]=
\frac{\effe{\f}\!\!_{+-}-\effe{\f}\!\!_{-+}-i(\effe{\f}\!\!_{h3}-\effe{\f}\!\!_{3h})  }{4}
\ea

\item   states 
$\Ttot=1,\Ytot=0$
\ba\label{10+1}
\ff{B3}{(1,0;+)}&\equiv& 
\F{}{B3}
\\\label{10+2}
\ff{\f}{(1,0;+)}&\equiv& {\rm Tr}[{\cal T}_L^3{\cal T}_R^3\F{}{\f}]
=\frac{\effe{\f}\!\!_{-+}+\effe{\f}\!\!_{+-}-\effe{\f}\!\!_{33}-
\,\effe{\f}\!\!_{hh}  }{4}\\
\label{10+3}
\ff{L_i}{(1,0;+)}&\equiv&\frac{1}{2} {\rm Tr}\left[t^3(\F{}{L_i}+\F{}{\bar{L_i}})\right]
=
\frac{ -\effe{L}\!\!_{ee}+\effe{L}\!\!_{\nu\nu}}{4}+
\frac{-\effe{L}\!\!_{\bar e \bar e}+
\effe{L}\!\!_{\bar \nu \bar \nu}}{4}
\\\nonumber
\\
\label{10-1}
\ff{L_i}{(1,0;-)}&\equiv&
 \frac{1}{2} {\rm Tr}\left[t^3(\F{}{L_i}-\F{}{\bar{L_i}})\right]
=
\frac{ -\effe{L}\!\!_{ee}+\effe{L}\!\!_{\nu\nu}}{4}-
\frac{-\effe{L}\!\!_{\bar e \bar e}+
\effe{L}\!\!_{\bar \nu \bar \nu}}{4}
\\
\label{10-2}
\ff{\f}{(1,0;-)}&\equiv&-\frac{1}{2} {\rm Tr}[{\cal T}_L^3\F{}{\f}]=
\frac{\effe{\f}\!\!_{+-}-\effe{\f}\!\!_{-+}+i(\effe{\f}\!\!_{h3}-\effe{\f}\!\!_{3h})  }{4}\\
\label{10-3}
f_g^{(1,0;-)}&\equiv&
-\frac{1}{2}\, {\rm Tr}[{ T}_L^3\F{}{g}]
= \frac{\effe{g}\!\!_{+-} - \effe{g}\!\!_{-+}}{2}
\ea
 
\item   states 
$\Ttot=2,\Ytot=0,\,CP=(+) $
\ba
\ff{g}{(2,0;+)}&\equiv&\frac{1}{4} 
{\rm Tr}[(3({T}_L^3)^2-2)\F{}{g}]=
\frac{\effe{g}\!\!_{+-}-2\,\effe{g}\!\!_{33}+\effe{g}\!\!_{-+}   }{4}
\ea

\item   states 
$ \Ttot=1,\Ytot^2=1,\,CP=(-) $
\ba
\ff{\f}{(1,1;-)}&\equiv& \frac{1}{2}{\rm Tr}[{\cal T}_y^3\,\F{}{\f}]=
\frac{\effe{\f}\!\!_{3h}+\effe{\f}\!\!_{h3}}{2}
\ea
\item   states 
$ \Ttot=1,\Ytot^2=1,\,CP=(+) $
\ba\label{fine}
\ff{\f}{(1,1;+)}&\equiv& \frac{i}{2}{\rm Tr}[({\cal T}_y^3{\cal T}_h^3)\,\F{}{\f}]=
\frac{\effe{\f}\!\!_{33}-\effe{\f}\!\!_{hh}}{2}
\ea

\item   states 
$\Ttot=1/2,\Ytot^2=\frac{1}{4} $

\be\label{lr}
\ff{LR_i}{\!\!_u}{(\frac{1}{2},\frac{1}{4};\pm)}
\equiv\frac{1}{2}\left(\,\effe{LR_i}\!\!_{11}\pm\effe{\bar{L}\bar{R}_i}\!\!_{11}\right)
\ee

\be\label{lbr}
\ff{LR_i}{\!\!_d}{(\frac{1}{2},\frac{1}{4};\pm)}
\equiv\frac{1}{2}\left(\,\effe{LR_i}\!\!_{22}\pm\effe{\bar{L}\bar{R}_i}\!\!_{22}\right)
\ee

\end{itemize}


\def\np#1#2#3{{\sl Nucl.~Phys.\/}~{\bf B#1} {(#2) #3}} \def\spj#1#2#3{{\sl
Sov.~Phys.~JETP\/}~{\bf #1} {(#2) #3}} \def\plb#1#2#3{{\sl Phys.~Lett.\/}~{\bf
B#1} {(#2) #3}} \def\pl#1#2#3{{\sl Phys.~Lett.\/}~{\bf #1} {(#2) #3}}
\def\prd#1#2#3{{\sl Phys.~Rev.\/}~{\bf D#1} {(#2) #3}} \def\pr#1#2#3{{\sl
Phys.~Rep.\/}~{\bf #1} {(#2) #3}} \def\epjc#1#2#3{{\sl Eur.~Phys.~J.\/}~{\bf
C#1} {(#2) #3}} \def\ijmp#1#2#3{{\sl Int.~J.~Mod.~Phys.\/}~{\bf A#1} {(#2) #3}}
\def\ptps#1#2#3{{\sl Prog.~Theor.~Phys.~Suppl.\/}~{\bf #1} {(#2) #3}}
\def\npps#1#2#3{{\sl Nucl.~Phys.~Proc.~Suppl.\/}~{\bf #1} {(#2) #3}}
\def\sjnp#1#2#3{{\sl Sov.~J.~Nucl.~Phys.\/}~{\bf #1} {(#2) #3}}
\def\hepph#1{{\sl hep--ph}/{#1}}

\end{document}